# Non-hermitian magnonic knobbing between electromagnetically induced reflection and transparancy

Youcai Han,[1][†] Changhao Meng,[1][†] Zejin Rao,[1][†] Jie Qian,[1] Yiming Lv,[1] Liping Zhu,[1] CanMing Hu,[2]* Zhenghua An,[1,3,4,5]*

*1 State Key Laboratory of Surface Physics, Institute of Nanoelectronic Devices and Quantum Computing, Department of Physics, Fudan University, Shanghai 200433, China*
*2 Department of Physics and Astronomy, University of Manitoba, Winnipeg R3T 2N2, Canada*
*3 Shanghai Qi Zhi Institute, 41st Floor, AI Tower, No. 701 Yunjin Road, Xuhui District, Shanghai, 200232, China*
*4Zhangjiang Fudan International Innovation Center, Fudan University, Shanghai 201210, China.*
*5Yiwu Research Institute of Fudan University, Chengbei Road, Yiwu City, 322000 Zhejiang, China*
†:these authors contribute equally to this work
*Email: Can-Ming.Hu@umanitoba.ca
*Email: anzhenghua@fudan.edu.cn

**Abstract**
Manipulation of wave propagation through open resonant systems has attracted tremendous interest. When accessible to the open system, the system under study is prone to tempering to out of equilibrium, and a lack of reciprocity is the rule rather than the exception. Open systems correspond to non-hermitian Hamiltonians with very unique properties such as resulting exceptional points and ideal isolation. Here, we have found a highly sensitive modulation for the intersection of resonant patch antennas with respect to cavity magnonic coupling by means of an open coupling system of three resonant modes. Two types of crossings are implemented in this study: the first type of crossing remotely controls the sharp switching of the transmission line 's transmittance, while regulating the repulsive behavior of its zero-reflection states. The second type of crossing corresponds to the modulation of non-reciprocal phase transitions, which enables a more desirable isolation effect. Three different coupling models are realized by a non-Hermitian scattering Hamiltonian, revealing distinct spatial overlaps between modes. This elucidates that dissipative coupling of at least two modes to the environment is crucial for non-reciprocal transport. Our work not only reveals the versatility of cavity magnonic systems but also provides a way to design functional devices for general wave optics using patch antenna crossings.

**Introduction**

Conservation of energy is a fundamental concept used to describe physical systems. In a closed system, the Hamiltonian of the system is Hermitian and its corresponding eigenvalues are purely real. Most of the time this is used to describe an overall total system, however there are times when we want to focus on one or some subsystems. In this case, the system of interest can exchange energy or matter between the particular system specified and the environment it is in[1-3]. This particular subsystem will become an open system. These systems have Hamiltonian whose eigenvalues will correspond to complex[4]. Their imaginary part represents the exchange of energy or matter between the subsystem and the external environment. The Hamiltonian of this subsystem which usually accompanied by non-reciprocal phase transitions[5,6] will be non-Hermitian. One of the most important of non-hermitian is the Parity-Time (PT) symmetry[1,7-9], which breaks the traditionally held belief that only hermitian Hamiltonian can exhibit purely real spectral behaviour. Non-hermitian systems also have unique behaviours such as non-hermitian skinning effects[10-15], fractional topological charges[16], complex energy band braiding[4,17], and loss-induced transparency[18]. Non-hermitian studies in the fields of acoustics[10,19-21], optics[1,22], heat transport[23], thermal atom[24,25] and optical mechanics[26,27] have been very colourful. Recent studies based on the cavity-magnonic coupling system have shown that the energy exchange between the transmission waveguide is considerable to the coherent coupling rate. Non-reciprocity transmission[28], ideal isolation[29], perfect microwave coherent absorption[30,31] have been achieved in the cavity-magnonic system.

The main focus has been on the study of their scattering spectra, and the study of their eigenvalues has concentrated on two modes. For the coupling of two resonant modes, the hybridised eigenstates exhibiting either level repulsive or level attractive[32-34] are relatively well defined. The behaviour of coupling due to three or more resonant modes is less clear. It was even found to exhibit energy level attraction in its reflection spectrum and level repulsion behaviour in its unit transmission[29] (UT) spectrum.

This study investigates the coupling dynamics of two interconnected cavities incorporating a ferromagnetic resonance mode. We observe distinct characteristic behaviors arising from varying coupling strengths among the three modes within the scattering spectrum. Furthermore, we identify contrasting behaviors in the corresponding eigenvalues across three distinct coupling scenarios. Notably, we identify two phase transition points delineating different coupling regimes. One set of phase transition points marks the transition from the zero-reflection state to total reflection, while the other signifies the shift from reciprocal to non-reciprocal transmission. These findings unveil the unique coupling dynamics among the three modes and offer novel insights for leveraging non-Hermitian cavity magnonic systems.

## Results and discussion
### Structure of the coupled cavity and magnon system
In the diagram of our device (Fig. 1a), a patch antenna consisting of three rectangular pieces of copper. There are two X waveguides and one Y waveguide, the longer X waveguide is impedance matched to the two ports and acts as a transmission line (TML), which subsequently named TML. The individual Y waveguide and TML form a T-shaped electromagnetic resonator. A single T-shaped electromagnetic resonator has a resonance frequency of 3.857 GHz, which corresponds to three-quarters of the wavelength and is a bright mode (B mode in Fig. 1b) with the line width of 314MHz as Fig.1c showed. Adding a quarter resonant wavelength X waveguide to the T-shaped electromagnetic resonator, a dark mode (D mode in Fig. 1b) as the same center frequency of 3.857 GHz can be found at the upper boundary of the Y waveguide and the right boundary of the X waveguide (Fig. 1b). B-mode and D-mode interactions exhibit rabi-like strong coupling, inducing a transparent window at the original resonance frequency (Fig. 1d). A 0.5 mm diameter yttrium iron garnet (YIG) single crystal sphere is placed on the surface of the patch antenna by means of a displacement stage, and an applied magnetic field (H-field) is applied perpendicular to the plane of the waveguide with a magnitude larger than the saturation field of the YIG sphere. Microwave signals are loaded from either port 1 or port 2 to excite ferromagnetic resonance (FMR) in the two YIG spheres. The frequency of the individual magnetic modes can be flexibly modulated by the applied magnetic field ($\omega_m \propto H$), and for simplicity the Kittle mode is generally chosen for the study. The magnon mode have a small line width of only 2 MHz (Fig. 1d) show substantially narrower resonances compared to the cavity, and have a large Q value equal to about 1000. The YIG sphere is moved in the y direction on the Y waveguide by a cantilevered rod with a displacement stage, which allows flexible modulation of the non-hermitian three-modes system.

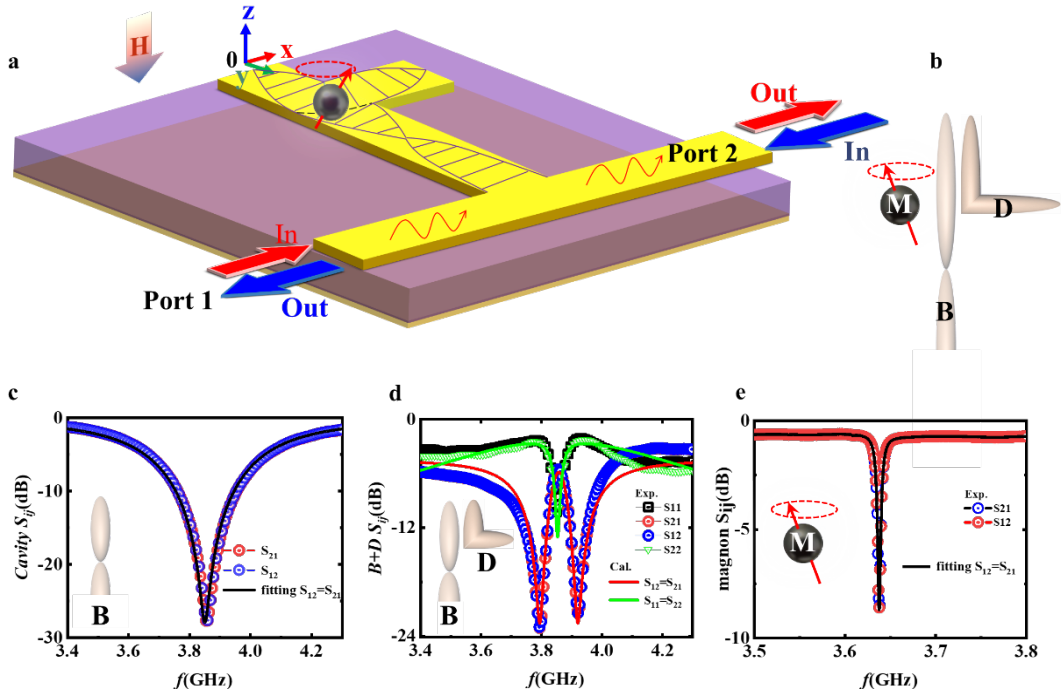

**Fig. 1 | Device structure and basic characteristics. a** Schematic diagram of the cavity magnonic sample. A Y-shaped cavity resonator, which contained a Y-waveguide that is three-quarters of the resonant wavelength and an X-waveguide that is one-quarter of the resonant wavelength, is coupled to the TML, concurrently supporting a bright (B) mode and a dark (D) mode. The purple envelope curve represents the magnetic field intensity distribution of the standing wave on the surface. A YIG crystal sphere is placed on the Y-shaped resonator. An externally applied magnetic field perpendicular to the sample plane allows flexible control of the magnetic field strength. The two ports of the TML are connected to a vector network analyzer to carry the traveling wave, where the red waveform represents the wave traveling from port 1 to port 2. **b** Schematic diagram illustrating the distribution of three modes (magnon mode (M), bright mode (B), and dark mode (D)) on the Y-shaped resonator. **c** Transmittance spectra of the uncoupled bright mode, obtained by measuring the transmission spectra (red for $|S_{21}|$, blue for $|S_{12}|$) when only the Y waveguide and TML are present(without the quarter-wavelength-long X waveguide). The black line represents the fitting curve. **d** Scattering spectra without the YIG crystal sphere (red for $|S_{21}|$, blue for $|S_{12}|$, green for $|S_{11}|$, black for $|S_{22}|$). The red line (black line) represents the calculated curve for transmission (reflection). **e** Transmission spectra of the YIG crystal sphere placed on a separate TML (red for $|S_{12}|$, black for $|S_{21}|$), with the black line indicating the fitted data. The resonance

frequency can be flexibly adjusted by the external magnetic field H.

**Non-hermitian three-mode coupled systems with three different coupling models**

For non-hermitian system, the approach of solving the quantum master equation[35,36] using Markov approximation is generally taken to study it. Time-domain coupled mode theory[37] is also a very good approach for this type of side-coupling waveguide system.

To describe such an open cavity magnonic system, We start with the effective resonant Hamiltonian of the open system. In general, for the uncoupled complex resonant frequency can be expressed as $\widetilde{\omega}_i = \omega_i - i\gamma_i$, where $i = b, d, m$ denote the cavity's bright and dark modes and magnon modes, and $\gamma_i$ corresponds to its intrinsic damping rate, respectively. The effective *resonant* Hamiltonian ($H_{res}$) taking account of the openness of the coupled system[37-39] can be written as,

$$H_{res} = H_0 - \frac{i}{2}D^\dagger D = \begin{bmatrix} \widetilde{\omega}_b - i\kappa_b & J_1 & J_2 \\ J_1 & \widetilde{\omega}_d & J_3 \\ J_2 & J_3 & \widetilde{\omega}_m - i\kappa_m \end{bmatrix} \quad (1)$$

In this configuration, the B-mode and magnon mode are coupled to the reservoir (TML) whit external damping rates of $\kappa_b$ and $\kappa_m$ rate, whereas the D-mode has no direct coupling to the TML($\kappa_d \cong 0$) due to lack of direct access to energy exchange. where $H_0 = \begin{bmatrix} \widetilde{\omega}_b & J_1 & J_2 \\ J_1 & \widetilde{\omega}_d & J_3 \\ J_2 & J_3 & \widetilde{\omega}_m \end{bmatrix}$ is the Hamiltonian of the closed system with coupled cavity magnon being isolated from the environment (MTL) except intrinsic damping ($\gamma_i$). $D = \begin{bmatrix} \sqrt{\kappa_b} & 0 & \sqrt{\kappa_m} \\ \sqrt{\kappa_b} & 0 & -\sqrt{\kappa_m} \end{bmatrix}$ is the coupling matrix between three resonant modes and two ports. The coherent coupling rate between the B-mode and D-mode is $J_1$, meanwhile $J_2$ and $J_3$ represent their respective coherent coupling rate to the magnon. The scattering-matrix ($S$-matrix) in the complex frequency is $S = C\left[I - iD\frac{1}{\omega - H_{res}}D^\dagger\right]$ (2) where $\omega$ is the complex frequency in the non-Hermitian scenario, $C$ is the coupling matrix between input and output ports and, in our present configuration, $C = \begin{bmatrix} 0 & 1 \\ 1 & 0 \end{bmatrix}$. $I$ is the unit matrix, i.e., $I = \begin{bmatrix} 1 & 0 \\ 0 & 1 \end{bmatrix}$. The effective resonant Hamiltonian $H_{res}$ had used to represent the eigenbehaviour of the scattering spectrum, it is found that $H_{res}$ describes the eigenbehaviour of the poles of the scattering spectrum when dealing with multi-mode coupled non-reciprocal open systems. In order to further describe its true eigenbehaviour at different scattering potentials, The effective *scattering* Hamiltonian $H_{sca}$ was raised, which can more accurately describe the scattering eigenbehaviour.

$$H_{sca}^{S_{i,j}=\lambda} = H_{res} - \frac{i}{\lambda - C_{i,j}}|\phi_{i,in}\rangle\langle\phi_{j,out}| \quad (3)$$

The input and output vectors are $|\phi_{1,in}\rangle = \begin{bmatrix} \sqrt{\kappa_b} & 0 & \sqrt{\kappa_m} \end{bmatrix}^T$, $|\phi_{2,in}\rangle = \begin{bmatrix} \sqrt{\kappa_b} & 0 & -\sqrt{\kappa_m} \end{bmatrix}^T$ and $\langle\phi_{1,out}| = \begin{bmatrix} \sqrt{\kappa_b} & 0 & -\sqrt{\kappa_m} \end{bmatrix}$, $\langle\phi_{2,out}| = \begin{bmatrix} \sqrt{\kappa_b} & 0 & \sqrt{\kappa_m} \end{bmatrix}$. (see Supplementary Text S2).

The physical advantage of $H_{sca}^{S_{i,j}=\lambda}$ is that the eigenvalues of $H_{sca}^{S_{i,j}=\lambda}$ can be seen directly from the

S spectra with a arbitrary target complex S-element value of $S_{i,j} = \lambda$. $H_{sca}$ shows that the scattering problem is an open system and that the dissipative coupling effects in the external part of the system directly affect the scattering spectrum.

We found different coupling conditions when the YIG were placed at different positions in the Y-waveguide, there are two unique phase transition points exist here, one at the intersection of the X-waveguide with the Y-waveguide and the other at the intersection of the Y-waveguide with the TML. As shown in the Fig. 2a-b at y = 0 mm and y = 36 correspond to phase transition points 1 and 2 respectively. Three different coupling regions (I:x=0, y= 0-12 mm, II:x=0, y= 12-36, III:x= -12-12 mm, y=36.5mm) corresponded to the YIG sphere as it moved from the very top of the Y waveguide to the TML. In region 1, the B-mode and D-mode overlap, and when the YIG is placed in this region, the three modes are directly coherently coupled to each other (Fig.2a). At this point, one of the intermediate hybridisation modes corresponds to a state of zero transmission (ZT) at the transmission frequency (indicated by the black arrow in Fig.2c). In regions I and II, the magnon mode is not in contact with the TML, so there is no direct coupling with the TML such that $\kappa_m = 0$. In region 2, the magnon mode no longer has any spatial overlap with the D-mode, such that $J_3$ goes straight to zero (Fig.2a). At the $\Delta = -5MHz$ the transmission has a additional peak corresponds to a UT state (indicated by the black arrow in Fig.2c). In region 3, the magnon begins to coupling directly with the TML where it produces a considerable $\kappa_m$. At this point the transmission spectral lines show a non-reciprocal behaviour, with $|S_{12}|$ showing a valley and $|S_{21}|$ showing a peak at the frequency indicated by the arrows in Fig.2c. We obtained the coupling coefficients for the YIG sphere at different positions in the Y-Waveguide by fitting Eq. (2) as shown in Fig. 2d. It can be found that the sudden change at the two crossing points corresponding to $J_3$ and $J_2$ respectively. These two mutation points also correspond to drastic changes in the scattering spectrum as shown in Fig. 2f, accompanied by a phase transition from zero to unit transmission. The changes in the spectral lines are extremely sensitive to the position of the YIG at these two turning points.

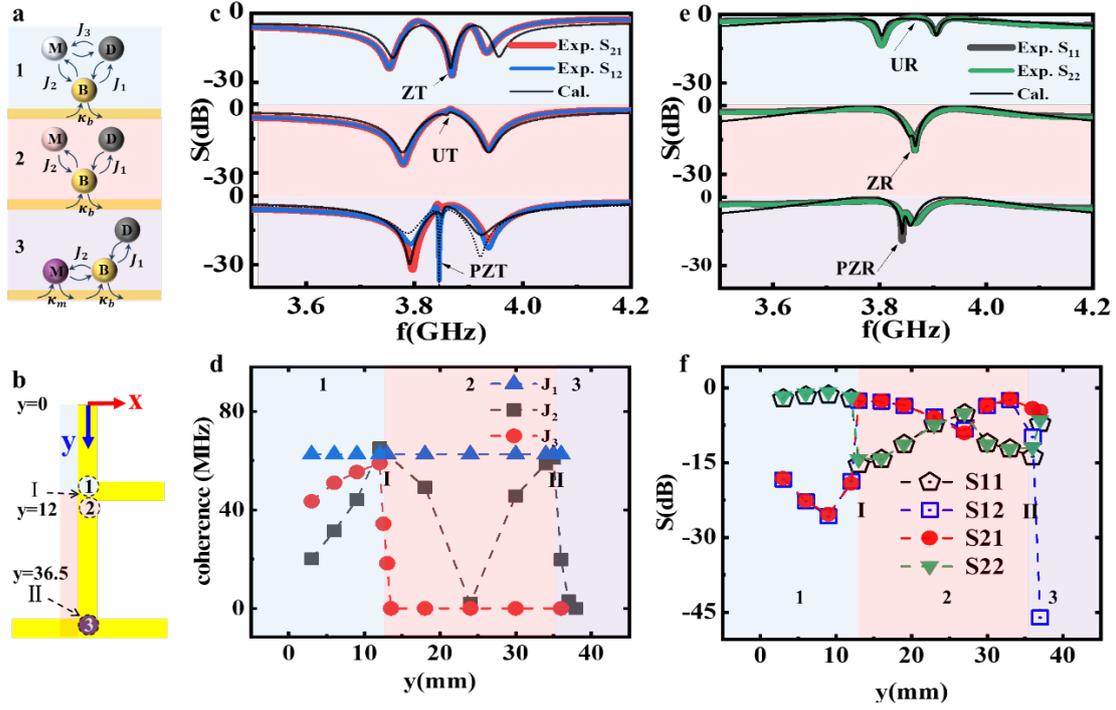

**Fig. 2 | Distinct scattering spectra transitions involving three coupling regions and two unique phase transition points. a** Schematic diagram illustrating coupling models in the three regions, represented by blue, peach, and purple colors for 1, 2, and 3, respectively. **b** A schematic diagram of the three coupling regions delineated by points I (y=12mm) and II (y=36mm), where the representative measuring points are 1 (y=12mm), 2 (y=12.5mm), and 3 (y=36.5mm). **c** The transmission spectra (red (blue) curves for $|S_{12}|$ ($|S_{21}|$)) in three coupling regions reveal distinctive resonant features induced by magnon introduction. From top to bottom, the curves correspond to the positions 1, 2, and 3, respectively, as illustrated in Figs. (a) and (b). the middle resonant frequency, indicated by the arrow, transitions from zero transmission (ZT) to unit transmission (UT) (1 to 2). From 2 to 3, only $|S_{12}|$ (red detuned ($\Delta_m$=-5MHz)) shifts to perfect zero transmission (PZT), by the way, $|S_{21}|$ shifts in the blue-detuned case, as shown in the supplementary materials. Solid black lines represent theoretical curves. In regions 1 and 2, within the reciprocal regime ($|S_{12}|=|S_{21}|$); in region 3, non-reciprocal phenomena emerge ($|S_{12}|\neq|S_{21}|$). **d** Relationship between the coherent coupling strength among the three modes and magnon position (y-coordinate values) obtained through theoretical formula fitting. Abrupt changes occur at positions I and II. **e** Reflection spectra corresponding to the three regions, where gray (green) lines depict $|S_{11}|$ ($|S_{22}|$), and black lines represent theoretically calculated curves ($|S_{11}|=|S_{22}|$). Moving the YIG crystal sphere illustrates zero-reflection undergoing Rabi-like strong coupling (1), decoupling (2), and dissipative coupling (3) phenomena in turn. **f** The variation of S-parameters with the magnon at different positions when $\Delta_m$=-5MHz is illustrated. In regions 1 and 2, two distinct phase transition behaviors are evident.

## The zero to unit transmission (Z2U) turning point

We now examine the nature of the Z2U turning point, which is the critical point of the coupling region 1 and 2. We set the $\kappa_m = 0$ because there is no direct coupling between the magnon and the TML. Figures 3a-f show, respectively, the mapping of $|S_{11}|$ or $|S_{21}|$, which are plotted as a function of the frequency detuning ($\Delta = \omega - \omega_b$) and field detuning ($\Delta_m = \omega_m - \omega_b$).

We focus on the uncoupling cavity mode $\omega_b = 3.857GHz$. From Fig. 3a to Fig. 3b, magnon moves only 2 mm on the Y waveguide, but achieves a ZT to UT switch at $\omega_b$ frequence.

We get the UT by setting the $S_{i,j} = 1$, in this condition, the eigenvalue of the Eq. (3), $\widetilde{\omega}_{UT} = \frac{1}{2}((\widetilde{\omega}_d + \widetilde{\omega}_m) \pm \sqrt{(\widetilde{\omega}_d - \widetilde{\omega}_m) - J_3^2})$, correspond to two high transmission hybridized states. When we set $S_{i,i} = 0$,

We obtain eigenvalues $\widetilde{\omega}_{ZR}$ of the same form as $\widetilde{\omega}_{UT}$, both of them showing anti-crossing in their spectral behaviour in region 1 and crossing in region 2 as Figs. 3a-b showed. When the detuning of the magnon and uncoupling B-mode is zero, i.e. $\Delta_m = \widetilde{\omega}_d - \widetilde{\omega}_m = 0$, the difference between the upper and lower branch $\Delta_\pm = \widetilde{\omega}_+ - \widetilde{\omega}_- = 2J_3$. This theoretically suggests that UT and zero reflection (ZR) behave in the same way and exhibit reciprocity, with both their level repulsions being dominated by by the non-zero $J_3$. (See Supplementary Text S2.1 for details) Fitting the experimental transmission scattering spectra at $\Delta_m = 0\ MHz$ using Eq. (3) yields $J_3=23\ MHz$ in region 1 and $J_3=0\ MHz$ in region 2, respectively. It is further shown that the $J_3$ value drops sharply to zero at this Z2U turning point leading to an anti-cross phase transition in the spectrum. We get the ZT by setting the $S_{i,j} = 0$, in this condition, the eigenvalue of the Eq. (3) is $\widetilde{\omega}_{ZT}$,

In Fig. 3e it can be seen that the eigenvalues in the middle of ZT intersect the real part of the eigen solution of UT at $\omega = \omega_c$, This is not a contradiction because their imaginary parts (Fig. 3h) are different, $\text{Im}(\omega_{ZT}) \neq \text{Im}(\omega_{UT})$, and therefore at this point it is the UT that is acting, and the value of the imaginary part of ZT is too large, and this state is decayed. Figure 3g shows that when the D-mode and magnon mode form a hybridised state in the region 1, the linewidths (imaginary parts) of their hybridised modes change accordingly, and at $\Delta_m = 0$, the imaginary parts of the two hybridised modes are equal $\text{Im}(\omega_{ZR+}) = \text{Im}(\omega_{ZR-})$.

This Z2U-point is the boundary point of the dark mode, and at the same time is the anti-node of the dark mode, At this juncture, it signifies the peak coherence in the coupling between the magnon and the dark mode (Fig. S5(b)), when leaving the Z2U -point and entering the 2 region, the magnon has no interaction with the D-mode due to the absence of spatial overlap, which also corresponds to the mutation of $J_3$. (Fig. S5(d)) Thus the presence of Z2U point can serve as a good position sensor.

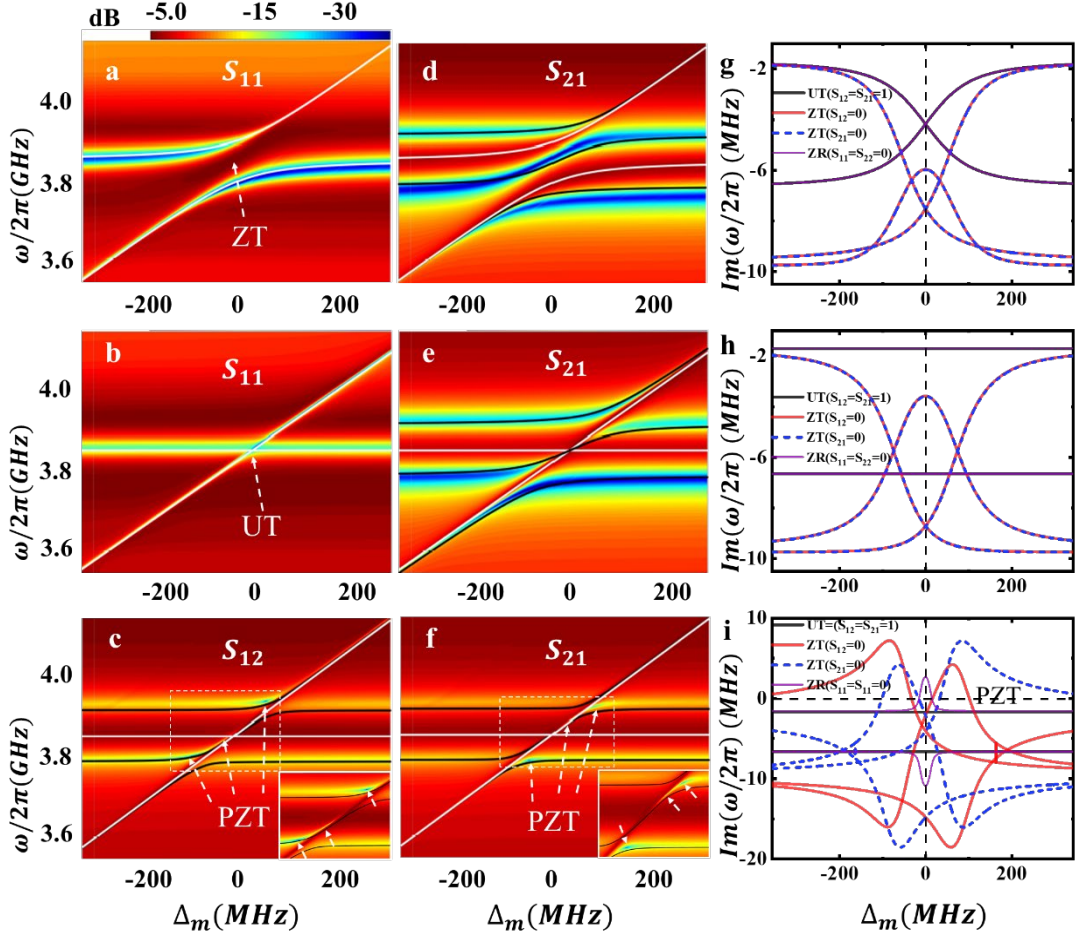

**Fig. 3 | Experimental S-spectra with scanning the M₂ detuning and the eigenvalues of the scattering Hamiltonians in three coupling regions.**

**a-f** Mappings of $|S_{11}|$(a-b),$|S_{12}|$(c), and $|S_{21}|$(d-f) as functions of $\omega$ and $\Delta_m$ in the three cases: the magnon in region 1(a, d), region 2(b, e), and region 3(e, f). In panels (a) and (b), the white line represents real zero reflection (ZR) eigenvalues. In panels (c) to (f), the white line indicates the real eigenvalues of the UT, while the black line represents the real eigenvalues of ZT. In panel (c) and (f), the arrows indicate the three respective PZT points in $|S_{12}|$ and $|S_{21}|$. **h-i** The image part of the UT(black solid line for $|S_{12}|=|S_{21}|=1$), ZT(red solid line for $|S_{12}|=0$, blue dash line for $|S_{21}|=0$),and ZR(solid purple line for $|S_{11}|=|S_{22}|=0$ ). PZTs emerges when the imaginary part of the transmission eigenvalues become zero.

## The non-reciprocal phase transition point

Next，we set out to characterise non-reciprocal behaviour, which requires the system to have both coherent and dissipative coupling. A non-reciprocal phase transition phenomenon occurs when the YIG blob moves from the Y waveguide into contact with the X TML (Fig. 2b). Introducing dissipative coupling not only leads to non-reciprocal phenomena, but also results in the emergence of peculiar resonant states in the transmission spectra, characterized by extremely narrow linewidths. In theory, under ideal conditions, these linewidths can be suppressed to zero, a state referred to as perfect zero transmission (PZT). In Figs. 3c and 3f, three PZT states are observed, each corresponding to narrow linewidths in both $|S_{12}|$ and $|S_{21}|$ occurring at distinct frequencies. Theoretically, this configuration offers optimal isolation. Similarly, almost zero linewidth states are observed in the reflection spectra, referred to as perfect zero reflection (PZR), as shown in Supplementary Text S2.2. The theoretical calculations reveal that these states correspond to zeros of the imaginary parts of the eigenvalues. Two PZR states can intersect with the middle PZR in the ZT spectrum, enabling unidirectional perfect absorption and reverse high transmittance (Fig. 3i). At the middle PZR frenquence, the transmission spectrum $|S_{12}|$ switches to zero in the external $\Delta_m = -5MHz$, while $|S_{21}|$ still maintains the full transmission behaviour. If switched to $\Delta_m = 5MHz$ the reciprocity is reversed. We call this the ideal isolation point which occurs when $\Delta_m = \pm 5MHz, \Delta = \omega - \omega_c = \pm 2.5MHz$.

Therefore the phase transition point is a non-reciprocal phase transition point, at this time the magnon is in direct contact with the TML, so the value of its $\kappa_m$ dissipative coupling to the TML starts to be non-negligible. The difference between the forward ($|S_{21}|$) and backward ($|S_{12}|$) transmission amplitudes is extracted in the decibel scale, we define its isolation ratio as Iso=log($|S_{21}|/|S_{12}|$). By solving the eigenequations of ZT, we can obtain that the imaginary part of this isolation point is zero. (Fig. 3i)

Next we fix the $\Delta_m$ to be in the red detuned state that enables the ideal isolation case, i.e., $\Delta_m = -5MHz$. Then we used a motor displacement table to move the YIG sphere from x=-15mm to x=15mm through the intersection y=36.5mm. According to Fig. S7, we find that the isolation is maximised when the magnon is at in the intersection x=0, meanwhile, $|S_{11}|=|S_{22}|$, indicating that it is a non-reciprocal symmetric network at this point. Deviating from this point, $|S_{11}|\neq|S_{22}|$, it becomes an asymmetric network.

Figures 4a and 4b show, respectively, the mapping of $|S_{11}|$ and $|S_{22}|$, which are plotted as a function of the frequency detuning ($\Delta = \omega - \omega_c$) and the position ($x$) of the YIG sphere on the TML, when the $\Delta_m = 5MHz$ is under the blue detuning. Meanwhile, Figures 4c and 4d demonstrates the same relation of mapping as the fig.4a and fig.4b when $\Delta_m = -5MHz$ is red detuned. In the mapping plot, two zero-reflection modes are observed, one characterized by high dissipation cavity-like mode, indicated by the black dashed lines in Fig. 4a-d. The other corresponds to a low-dissipation mode. Using the cavity-like mode as a unified reference center frequency, it is observed that under the same $\Delta_m$, the frequency difference between the measured $|S_{11}|$ and $|S_{22}|$ is different, indicating asymmetry in the network.

So we define a asymmetric ratio Asy=log($|S_{11}|/|S_{22}|$) in the low-dissipation mode frenquence. When the YIG sphere is moved to x=-12mm, the Asy reaches its maximum Asy=-80dB, as shown in Fig. 4e. When moved to x=+12mm, its Asy is minus max. It is noteworthy that when the YIG crystal sphere is positioned at x=0mm, the network exhibits symmetry, yet its non-reciprocity reaches a maximum (maximum Iso value showed as Figs. S7(b) and S7(e)). This can be explained by examining the imaginary part of the eigenvalues of the zero-reflection states. Figure 4f presents the theoretical calculations of the imaginary parts of the low-loss mode corresponding to $|S_{11}|$ ($\omega_{11}$) and $|S_{22}|$ ($\omega_{11}$) spectra (when the $\Delta_m = -5$MHz and $\Delta_m = -5MHz$, the $Im(\omega_{ii}) = Im(\omega_{ii})$). The theoretical calculations are detailed in Supplementary Text S4. The black dashed box in Fig. 4f is the range of positions in Figs. 4a-d, where theoretical calculations show that the imaginary part of their eigenvalues is equal to zero at x = -12 and 12 mm, respectively, corresponding to the maximum Asy. It is shown that there is a certain symmetry here, and when the sign reversal of the $\Delta_m$ at the same time as the magnon mode position does the reflection operation, the spectrograms are inter-convertible between Fig. 4a and Fig. 4d (and at the same time Fig. 4b and Fig. 4c).

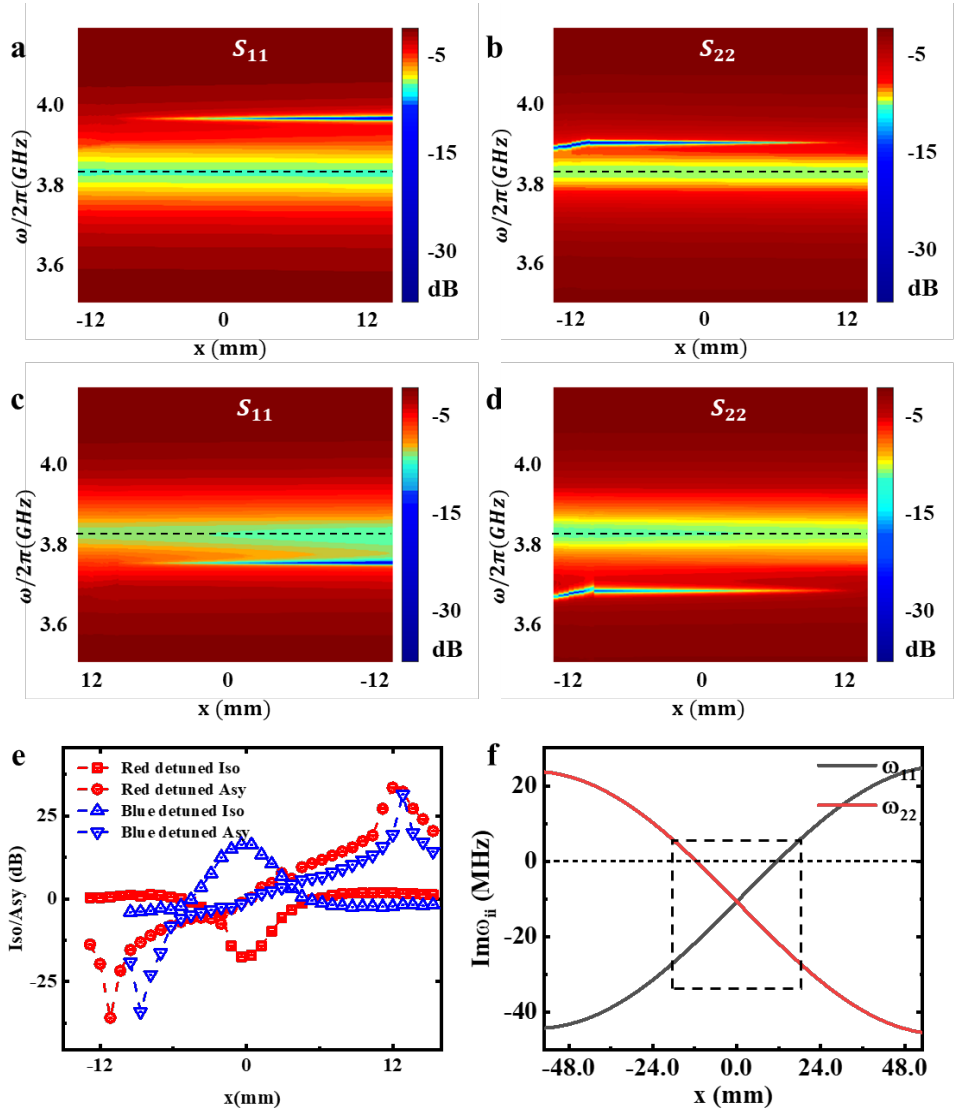

**Fig. 4 | Near phase transition point II: adjustment of asymmetric network**

**a-b** The $|S_{11}|$ (a) and $|S_{22}|$ (b) mapping as a function of $\omega$ and x at the magnon blue-detuned isolation point ($\Delta_m = 5MHz$). Two resonances appear, with the higher frequency having a smaller linewidth, indicating a low-loss mode that causes a split in $|S_{11}|$ and $|S_{22}|$. **c-d** The $|S_{11}|$ (c) and $|S_{22}|$ (d) mapping as a function of $\omega$ and x at the red-detuned isolation point ($\Delta_m = -5MHz$). The black dashed lines represent the cavity-like modes, serving as reference frequencies. The red detuning of $|S_{11}|$ equals the blue detuning of $|S_{22}|$ ((a) vs (d) or (c) vs (b)). **e** The Asy and Iso varies with the x-coordinate at the blue (red) $\Delta_m$ detuning. The Iso (Asy) is symmetrical (antisymmetrical) with respect to the X-coordinate. **f** The imaginary part of the low-loss mode of ZR$|S_{11}|$=0 (ZR$|S_{22}|$=0) versus position x calculated by Eq. (S21). At x=-5mm (x=5mm), the appearance of PZR corresponds to the maximum value of the Asy.

## conclusion

We have designed a three-mode cavity-magnetic coupling system and discovered two special types of waveguide turning point with sensitive modulation of microwave transmission.

In mode coupling, we find that the dark modes interacting with the magnon modes can affect the level repulsive behaviour of the reflection spectrum. At the same time due to the drastic changes in the scattering spectrum leaving the overlapping region of the dark modes, at the same time both can be characterised by spectral characterisation of the spatial distribution of the dark modes. It is also expected to be used as a position detector sensor, etc. Due to the Markov approximation[36], it can be assumed that the coherent coupling has a longer lifetime. The combined effect of dissipative coupling and coherent coupling leads to the phenomenon of non-reciprocal transmission. It is also shown that the field strength distribution on the patch antenna at the crossing point may have some special distribution. This work reveals that not only can the value of the transmission and reflection be sensitively regulated, but also the non-reciprocity as well as the asymmetry of the microwave network can be adjusted. The coupling parameters can be adjusted to a large extent over small distances, which provides a rather convenient experimental scheme for experiments on multi-resonator coupling.

## Materials and Methods

**The Device description.** The specimen is crafted on a 0.762 mm thick RO4350B substrate, which forms a square with a side length of 60 mm. The Y waveguide has a width of 1.686 mm and a length of 36 mm. The X waveguide has the same width but a length of 12mm, corresponding to a quarter of the wavelength of the uncoupled cavity mode. The X waveguide is connected at the upper quarter resonant wavelength of the Y waveguide. The complete cavity is enveloped by a 0.035 mm thick copper layer, ensuring impedance matching to 50 ohms. Two coaxial twin cables, affixed with MSW connectors at both ends, are employed for convenient input and output of microwave power to and from the sample. The resonant frequencies of the uncoupled bright and dark modes of the cavity are both 3.857 GHz, the diameter of the YIG single-crystal sphere is 0.5 mm, and the theoretical saturation magnetization intensity is below 0.2 T.

**Measurement setup.** An external magnetic field applied perpendicular to the cavity plane is utilized to precisely tune the detuning of magnon ($\Delta_m \equiv \omega_m - \omega_b$) by finely adjusting the ferromagnetic resonance frequency. Two ports are employed for obtaining S-spectra, encompassing both reflection and transmission coefficients. The reflection and transmission spectra are measured utilizing a vector network analyzer (VNA) operating with an input power level set at -5 dBm.

## Acknowledgments

Z.A. acknowledges the financial support from the National Natural Science Foundation of China under Grant Nos. 12027805/11991060/11634012/11674070 and the Shanghai Science and Technology Committee under Grant Nos. 18JC1420402, 20JC1414700, 20DZ1100604. C.-M.H. acknowledges the financial support from NSERC Discovery Grants and NSERC Discovery Accelerator Supplements. We acknowledge Doc. Huanyi Xue for guidance on drawing and ChatGPT 3.5 for proofreading. Part of the experimental work was conducted in Fudan Nanofabrication Lab.

**Author contributions:** Y.C.H. initiated the project and designed the three-resonator system. Y.C.H. finished all the experiments and discovered the phenomenon with the support of J.Q, Y.M.L, and Z.J.R. Theoretical framework was established by C.H.M. together with Z.A. and Y.C.H. Y.C.H. completed numerical analyses and calculations as well as visualization. Z.A. wrote the manuscript with Y.C.H, C.H.M, and J.Q. All authors commented on the manuscript. Z.A. and C.-M.H. supervised the whole project.

**Competing interests:** The authors declare no competing interests.

**Data and materials availability:** All data are available in the main text or the supplementary materials.

**Code availability:** All of the code used to support numerical calculations and plots are available from the corresponding author upon reasonable request.

**Supplementary material**

**S1. Non-hermitian theoretical model and setup**

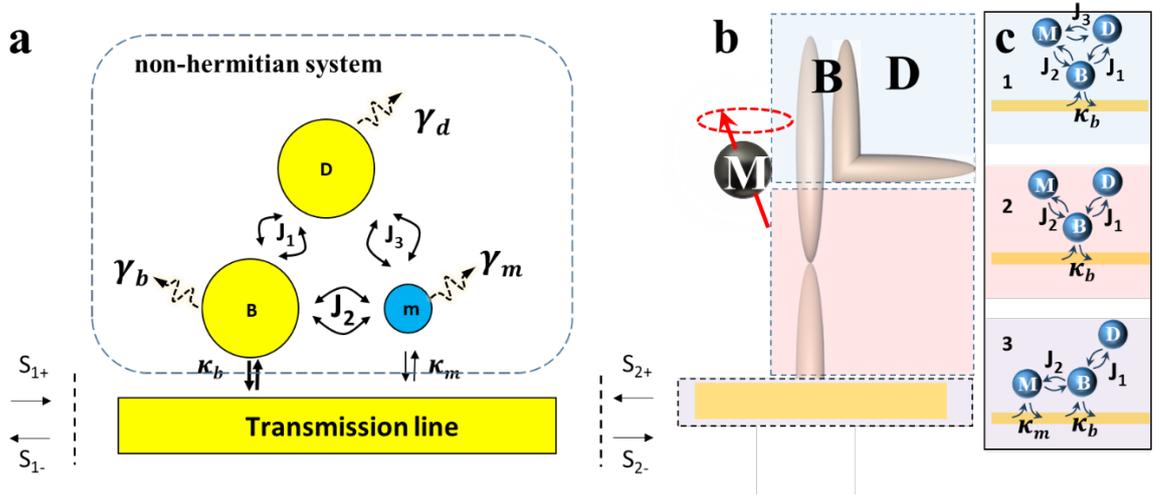

**FIG.S1 Model diagram depicting the coupling of non-Hermitian three modes.** (a) Schematic of simplified theoretical model of the three coupled resonator with a standard transmission line (MTL). (b) Spatial distribution of the three modes. (c) Model diagrams depicting coupling behaviors in three corresponding regions: Region 1 in blue (y=0-12mm), Region 2 in peach (y=12-36mm), and Region 3 in purple (y= 36-37.75mm).

Figure S1(a) shows the schematic model of our non-hermitian theoretical model. The bright mode is side-coupled to the transmission line (TML). When the YIG crystalline sphere approaches and comes into contact with the TML, magnon also side-coupled to the reservoir. Through the external damping rate ($\kappa_b$) of the bright mode or magnon mode ($\kappa_m$), both can exchange energy directly with the TML. There is a strong coupling behavior similar to Rabi between two bright and dark modes with the same resonant frequency supported simultaneously on the patch antenna, and their coupling strength is $J_1$. The $J_2$ and $J_3$ is the coherent coupling rate of the magnon mode with bright and dark mode, respectively. The uncoupled resonant complex frequencies are expressed as $\widetilde{\omega}_i = \omega_i - i\gamma_i$, where i=b,d,m correspond to the bright mode, dark mode, magnon mode respectively and $\gamma_i$ is the intrinsic damping rate.

Excitation of the resonant modes occurs through input waves $S_{1+}$ and $S_{2+}$ from ports 1 and 2, respectively. The resonant modes exhibit leakage due to the outgoing waves $S_{1-}$ and $S_{2-}$.

We use a two port vector network analyzer to measure the transmission and reflection spectra of the system. The external magnetic field is perpendicular to the circuit plane and can flexibly adjust the frequency of ferromagnetic resonance $\omega_m$. The specimen is fabricated on a 0.762 mm thick RO4350B substrate, which is in the shape of a square with a side length of 60 mm. The width of the Y-waveguide is 1.686 mm and the length is 36 mm, on the other hand, the X waveguide has the same width but a length of 12mm, corresponding to a quarter of the wavelength of the uncoupled cavity mode. The X waveguide is connected at the upper quarter resonant wavelength of the Y waveguide. The cavity is made of 0.035 mm thick copper with an impedance of 50 ohms. Two coaxial twin cables are connected at both ends to facilitate the microwave input and output. The uncoupled bright and dark cavity modes are all 3.857 GHz, which are determined by the corresponding designed waveguide size. We are using YIG crystal spheres with a radius of 0.5 millimeters and a saturation magnetization of 0.2 T. The magnetic field applied is greater than the range above its saturation field to regulate its ferromagnetic resonance frequency.

We focus on the uncoupled cavity mode frequencies $\omega_b$ and define the detuning between the magnon and uncoupled cavity mode as $\Delta_m = \omega_m - \omega_b$. We can use a cantilever to move the YIG crystal sphere along the y-axis centerline on the Y-waveguide. Because there will be different coupling behaviors when moving to different positions, it can be roughly divided into three regions. After fixing the position of YIG, we adjust $\Delta_m$ by applying the applied magnetic field H. This is thanks to the flexible and controllable ferromagnetic resonance.

Firstly, we design a resonant cavity with only Y waveguide and TML, and obtain the resonant frequency $\omega_b$ and linewidth $\gamma_b$ of the bright mode, and then obtain its coupling rate $\kappa_b$ with the TML by fitting. Subsequently, we obtain the intrinsic loss $\gamma_d$ of the dark mode and the strength of coherent coupling between the two modes $J_1$ by fitting the scattering parameter of the sample with both Y and X waveguides. Then we add the YIG crystal sphere to interact with the cavity modes, and their scattering elements are fitted to obtain the external loss $\kappa_m$ and intrinsic loss $\gamma_m$ of the magnon mode and its coherent coupling coefficients with the two cavity modes $J_2$, $J_3$ using the theoretical formulas presented later. The coupling coefficients obtained by numerical fitting is shown in Table S1.

## S2 Scattering Hamition

We optimize a theoretical framework employing the temporal coupled-mode theory[37-39] (TCMT) and derive effective non-Hermitian Hamiltonians for our open magnonic system. The Hamiltonian $H_0$ represents the coupled system of three resonators in the absence of external interactions.

$$H_0 = \begin{bmatrix} \widetilde{\omega}_b & J_1 & J_2 \\ J_1 & \widetilde{\omega}_d & J_3 \\ J_2 & J_3 & \widetilde{\omega}_m \end{bmatrix}, \qquad (S1)$$

The $\widetilde{\omega}_i = \omega_i - i\gamma_i$ where $i = b, d, m$, represent the resonant frequencies of the uncoupled bright, dark cavity mode and the magnon mode, respectively, with intrinsic loss rates $\gamma_i$. The J1 is the coherent coupling rate of the bright and dark modes, and the J2(J3) is the coherent coupling rate of bright (dark) mode and magnon. The Hamiltonian $H_0$ delineates the interplay among the three leaky resonant modes, temporarily neglecting the influence from the TML bus. The coupling channels to the TML bus can be modelled by a coupling matrix $D$.

$$D = \begin{bmatrix} \sqrt{\kappa_b} & 0 & \sqrt{\kappa_m} \\ \sqrt{\kappa_b} & 0 & -\sqrt{\kappa_m} \end{bmatrix}, \tag{S2}$$

The system's resonant Hamiltonian, incorporating optical dissipation in the coupling channels, can be expressed as $H_{res}$.

$$H_{res} = H_0 - \frac{i}{2}D^\dagger D = \begin{bmatrix} \widetilde{\omega}_b - i\kappa_b & J_1 & J_2 \\ J_1 & \widetilde{\omega}_d & J_3 \\ J_2 & J_3 & \widetilde{\omega}_m - i\kappa_m \end{bmatrix} \tag{S3}$$

The eigenvalues of $H_{res}$ are influential in dictating the resonances of the system, manifesting as poles in the complex frequency plane of the scattering (S) matrix. Utilizing the temporal coupled-mode theory (TCMT), the scattering (S) matrix can be deduced and expressed as

$$S = C\left[I - iD\frac{1}{\omega - H_{res}}D^\dagger\right] \tag{S4}$$

where $C = \begin{bmatrix} 0 & 1 \\ 1 & 0 \end{bmatrix}$ is the direct coupling matrix between the ports.

The effective scattering Hamiltonian, capable of characterizing arbitrary values $\lambda$ for any element within the scattering (S) matrix, is denoted as

$$H_{sca}^{S_{i,j}=\lambda} = H_{res} - \frac{i}{\lambda - C_{i,j}}|\phi_{i,in}\rangle\langle\phi_{j,out}| \tag{S5}$$

Where $|\phi_{\frac{1}{2},in}\rangle = \begin{bmatrix} \sqrt{\kappa_b} & 0 & \pm\sqrt{\kappa_m} \end{bmatrix}^T$ and $\langle\phi_{\frac{1}{2},out}| = \begin{bmatrix} \sqrt{\kappa_b} & 0 & \mp\sqrt{\kappa_m} \end{bmatrix}$ represent the input and output vectors characterizing the coupling between the input/output channels and the cavity-magnonic coupled system. These vectors can be derived from the D matrix in accordance with

$$[|\phi_{1,in}\rangle \;|\phi_{2,in}\rangle] = D^\dagger = \begin{bmatrix} \sqrt{\kappa_b} & \sqrt{\kappa_b} \\ 0 & 0 \\ \sqrt{\kappa_m} & -\sqrt{\kappa_m} \end{bmatrix} \text{ and } \begin{bmatrix} \langle\phi_{1,out}| \\ \langle\phi_{2,out}| \end{bmatrix} = CD = \begin{bmatrix} \sqrt{\kappa_b} & 0 & -\sqrt{\kappa_m} \\ \sqrt{\kappa_b} & 0 & \sqrt{\kappa_m} \end{bmatrix}.$$

We emphasize that our effective Hamiltonian has the capability to transform arbitrary scattering potential values $\lambda$ into eigenvalue problems by specifying the desired element of the S-matrix. In a broader context, we can systematically obtain the poles and zeros of the S-matrix, elucidating the conditions for resonance, absorption, and unidirectional transmission exceptional points (EPs).

## S2.1 The reciprocal zone

### zero-reflection and unit transmission.

For this resonant cavity structure, we designate the center of the intersection between the Y waveguide and the X waveguide as crossing point I, corresponding to y = 12 mm. The center of the intersection between the Y waveguide and the TML is marked as crossing point II, with the corresponding y = 36.5 mm, as illustrated in Fig. S1(b). When the YIG crystal sphere is positioned near crossing point I, the distant location of the YIG crystal sphere from the TML ensures that there is no wave function overlap, resulting in $\kappa_m = 0$.

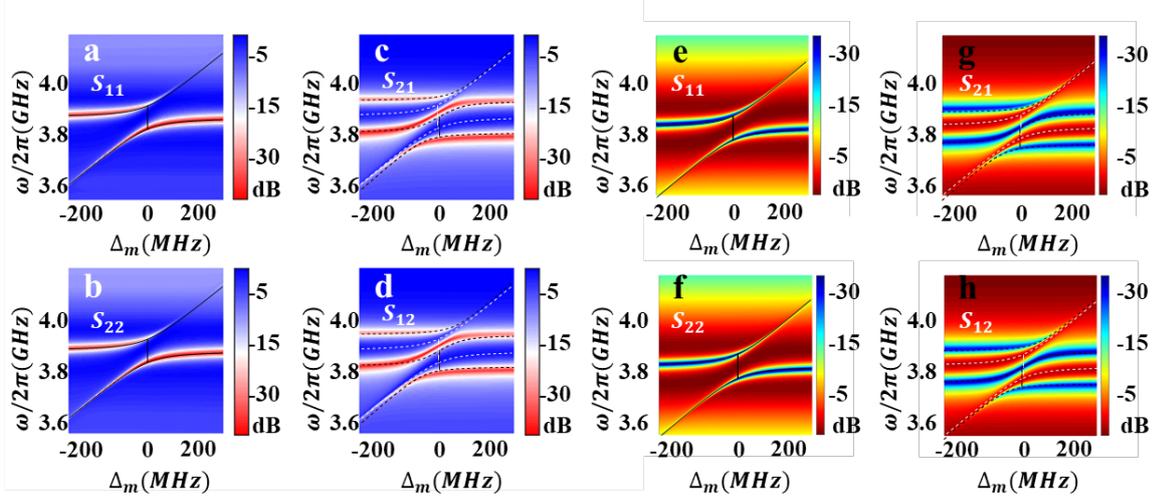

**FIG.S2 Scattering spectra and eigenvalue behaviors of the scattering Hamiltonian in Region 1.** (a)-(d) Experimental obtained spectra of four elements of the S-matrix as the function of the $\omega$ and the $\Delta_m$. (e)-(f) Calculated mapping of $|S_{11}|$ (e), $|S_{22}|$(f), $|S_{21}|$ (g), $|S_{12}|$ (h) as the function of the $\omega$ and the $\Delta_m$. The curves represent the eigenvalues of the scattering Hamiltonian. The solid (dash) black line represents ZR (ZT), and the white dashed line represents UT by theory calculation. ZR and UT exhibit a characteristic level repulsion (anti-cross).

When moving around this crossing point, there is a discontinuity in the zero reflection state and unit transmission state. The eigenvalue behavior of the scattering spectrum changes from anti-crossing to crossing over a short distance, resulting in a transition from zero reflection (unit transmission) to zero transmission at the central frequency. Therefore, our focus here is on the eigenvalue behavior of $S_{i,i} = 0$ and $S_{i,j} = 1$.

The effective scattering Hamiltonian $H_{sca}^{S_{i,i}=0}$ for $S_{1,1}, S_{2,2} = 0$ derived from Eq.(S5) can be written as

$$H_{sca}^{S_{i,i}=0} = \lim_{\epsilon \to 0+} \begin{bmatrix} \widetilde{\omega}_b - i\kappa_b + \frac{i\kappa_b}{\epsilon} & J_1 & J_2 \\ J_1 & \widetilde{\omega}_d & J_3 \\ J_2 & J_3 & \widetilde{\omega}_m \end{bmatrix} \quad (S6)$$

$$\det\left|\widetilde{\omega}_{ZR} - H_{sca}^{S_{i,i}=0}\right| = 0$$

$$(\widetilde{\omega}_b - i\kappa_b + \frac{i\kappa_b}{\epsilon} - \widetilde{\omega}_{ZR})((\widetilde{\omega}_d - \widetilde{\omega}_{ZR})(\widetilde{\omega}_m - \widetilde{\omega}_{ZR}) - J_3^2) - J_1^2(\widetilde{\omega}_m - \widetilde{\omega}_{ZR}) - J_2^2(\widetilde{\omega}_d - \widetilde{\omega}_{ZR})$$
$$+ 2J_1J_2J_3 = 0$$

$$\frac{i\kappa_c}{\epsilon}\left((\tilde{\omega}_d - \tilde{\omega}_{ZR})(\tilde{\omega}_m - \tilde{\omega}_{ZR}) - J_3^2\right) = 0$$

Then we get the zero reflection eigenvalues

$$\tilde{\omega}_{ZR} = \frac{1}{2}\left((\tilde{\omega}_d + \tilde{\omega}_m) \pm \sqrt{(\tilde{\omega}_d - \tilde{\omega}_m) - J_3^2}\right) \tag{S7}$$

When we set $S_{i,j} = 1$, we observe that $H_{sca}^{S_{i,j}=1}$ equals $H_{sca}^{S_{i,i}=0}$. Therefore, the eigenvalues of the unit transmission state are the same as those of the zero reflection state.

$$\tilde{\omega}_{UT} = \tilde{\omega}_{ZR} = \frac{1}{2}\left((\tilde{\omega}_d + \tilde{\omega}_m) \pm \sqrt{(\tilde{\omega}_d - \tilde{\omega}_m) - J_3^2}\right) \tag{S8}$$

A surprising result is that the bright mode has no discernible impact on the eigenvalue behavior of its unit transmission state and zero reflection state. It acts as a mere mediator, transferring the effects of dark modes and magnon modes onto the TML.

Therefore, we obtain two eigenfrequencies primarily determined by the magnitude of $J_3$. In the case where $J_3$ is nonzero, the eigenvalues of ZR (zero reflection) and UT (unit transmission) exhibit level repulsion behavior. Under the condition of $\Delta_m$, the difference between the two eigen-solutions, denoted as $\Delta = \omega - \omega_b$, is equal to $2J_3$.

To further investigate its nonreciprocal behavior, solving for the effective scattering matrix $H_{sca}^{S_{i,j}=0}$ corresponding to $S_{1,2} = 0$ and $S_{2,1} = 0$, nevertheless, we obtain the same $H_{sca}^{S_{i,j}=0}$.

$$H_{sca}^{S_{i,j}=0} = \begin{bmatrix} \tilde{\omega}_b & J_1 & J_2 \\ J_1 & \tilde{\omega}_d & J_3 \\ J_2 & J_3 & \tilde{\omega}_m \end{bmatrix}$$

Therefore, the behaviors of $S_{1,2}$ and $S_{2,1}$ are identical, indicating that this is a reciprocal system. Simultaneously, since the behaviors of $S_{1,1}$ and $S_{2,2}$ are also the same, it belongs to a symmetric microwave electronic network. Figs. S4a-d display the experimental scattering spectra, while Figs. S4e-h depict the corresponding theoretical mapping calculated. The solid (dash) black line represents ZR (ZT), and the white dashed line represents UT by theory calculation. ZR and UT exhibit a characteristic level repulsion (anti-cross). The experimental results agree well with the theoretical calculations.

As only the bright mode directly interacts with the TML, dissipative coupling behavior does not occur. Dissipative coupling behavior should involve at least two modes directly interacting with the external bus to take place[28]. In contrast, nonreciprocal behavior requires modes to simultaneously possess both coherent and dissipative coupling. This aligns well with both theoretical predictions and experimental results.Figures S3 shows the scattering spectra corresponding to region 2. It is observed that the behavior transitions to UT at $\Delta_m = 0$ and the center frequency $\omega$=3.857 GHz.

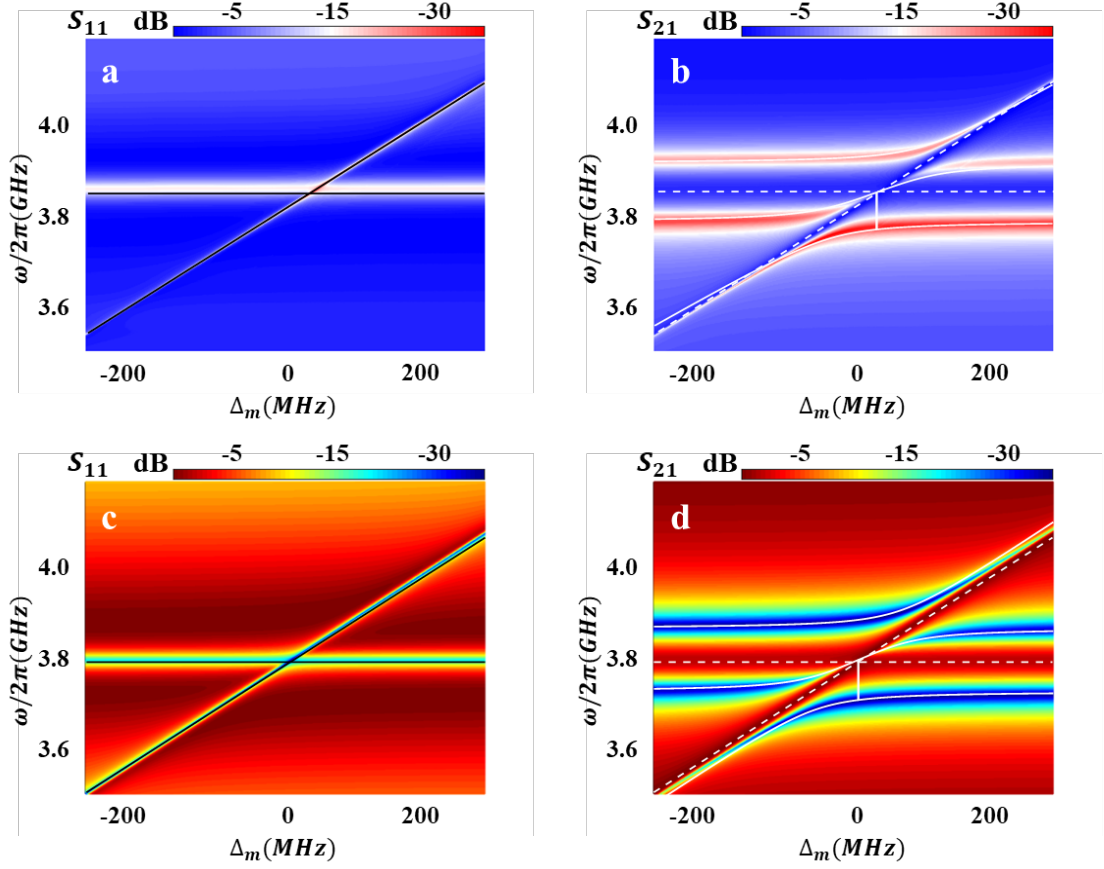

**FIG.S3 Experimental (top rows) and calculated (bottom rows) obtained spectra of four elements of the S-matrix as the function of the $\omega$ and the $\Delta_m$ in Region 2.** (a)-(b) Mapping of $|S_{11}|$ (a) and $|S_{21}|$ (b) spectra in experiment. (c)-(d) Mapping of $|S_{11}|$ (c) and $|S_{21}|$ (d) spectra under theoretical calculation. The solid black curves indicate the real part of the eigenvalues of the ZR. The ZR (UT) directly exhibits a crossing behavior.

### S2.2 The non-reciprocal zone

When the YIG crystal sphere is positioned near crossing point 2, the magnon is directly coupled to the TML as the YIG crystal sphere is in direct contact with it. In this scenario, Km cannot be neglected. However, it is evident that there is no coherent interaction with the dark mode at this point, resulting in $J_3$ being equal to zero.

From this, we obtain

$$H_{sca}^{S_{i,i}=0} = \lim_{\epsilon \to 0} \begin{bmatrix} \widetilde{\omega}_b - i\kappa_b + \dfrac{i\kappa_b}{\epsilon} & J_1 & J_2 \mp \dfrac{i\sqrt{\kappa_b \kappa_d}}{\epsilon} \\ J_1 & \widetilde{\omega}_d & 0 \\ J_2 \pm \dfrac{i\sqrt{\kappa_b \kappa_m}}{\epsilon} & 0 & \widetilde{\omega}_m - i\kappa_m - \dfrac{i\kappa_m}{\epsilon} \end{bmatrix}$$

The two positive and negative signs correspond to i=1 and i=2, respectively. It is known that $H_{sca}^{S_{1,1}=0} = (H_{sca}^{S_{2,2}=0})^T$, so their respective eigenvalues are also equal.

zero-reflection condition is given by,

$$det\left|\widetilde{\omega}_{ZR} - H_{sca}^{S_{i,i}=0}\right| = 0.$$

$$(\widetilde{\omega}_{ZR} - \widetilde{\omega}_d)[(\kappa_m - \kappa_b)\widetilde{\omega}_{ZR} + \kappa_b\widetilde{\omega}_m - \kappa_m\widetilde{\omega}_b] = \kappa_m J_1^2$$

Which is transformed into a quadratic equation of $\widetilde{\omega}_{ZR}$

$$(\kappa_m - \kappa_b)\widetilde{\omega}_{ZR}^2 + (\kappa_b\widetilde{\omega}_m - \kappa_m\widetilde{\omega}_b - (\kappa_m - \kappa_b)\widetilde{\omega}_d)\widetilde{\omega}_{ZR} + \kappa_m\widetilde{\omega}_d\widetilde{\omega}_b - \kappa_b\widetilde{\omega}_d\widetilde{\omega}_m - \kappa_m J_1^2 = 0 \quad (S9)$$

$$\widetilde{\omega}_{ZR\pm} = \frac{-B \pm \sqrt{B^2 - 4AC}}{2A} \quad (S10)$$

Where $A = \kappa_m - \kappa_b$, $B = \kappa_b\widetilde{\omega}_d - \kappa_d\widetilde{\omega}_b - (\kappa_m - \kappa_b)\widetilde{\omega}_d$, $C = \kappa_m\widetilde{\omega}_d\widetilde{\omega}_b - \kappa_b\widetilde{\omega}_d\widetilde{\omega}_m - \kappa_m J_1^2$.

We observed that the zero-reflection state exhibits a behavior of level attraction. When we set the real parts of the two branches of the zero-reflection state equal, we obtain an effective dissipation coupling strength denoted as

$$-i\sqrt{\frac{\kappa_m}{\kappa_b - \kappa_m}} J_1$$

**The unit-transparency condition**

Setting the target off-diagonal elements ($S_{i,j}$) of the $S$-matrix to be unit ($\lambda = 1$), we get the effective scattering Hamiltonian for Unit-transparency,

$$H_{sca}^{S_{i,j}=1} = \lim_{\epsilon \to 0}\begin{bmatrix} \widetilde{\omega}_b - i\kappa_b + \frac{i\kappa_b}{\epsilon} & J_1 & J_2 \pm \frac{i\sqrt{\kappa_b\kappa_m}}{\epsilon} \\ J_1 & \widetilde{\omega}_d & 0 \\ J_2 \pm \frac{i\sqrt{\kappa_b\kappa_m}}{\epsilon} & 0 & \widetilde{\omega}_m - i\kappa_m + \frac{i\kappa_m}{\epsilon} \end{bmatrix} \quad (S11)$$

Unit-transmission ($S_{21}=1$ or $S_{12}=1$) is given by,

$$det\left|\widetilde{\omega}_{UT} - H_{sca}^{S_{i,j}=1}\right| = 0$$

**The transmission bound states in the continuum (t-BIC) condition**

When setting the non-diagonal elements ($S_{i,j}$) of the S matrix equal to zero ($\lambda = 0$), we get the effective scattering Hamiltonian for zero-transmission,

$$H_{sca}^{S_{i,j}=0} = \begin{bmatrix} \widetilde{\omega}_b & J_1 & J_2 \pm i\sqrt{\kappa_b\kappa_2} \\ J_1 & \widetilde{\omega}_d & 0 \\ J_2 \pm i\sqrt{\kappa_b\kappa_m} & 0 & \widetilde{\omega}_m \end{bmatrix} \quad (S12)$$

zero-transmission condition is given by,

$$det\left|\widetilde{\omega}_{ZT} - H_{sca}^{S_{i,j}=0}\right| = 0$$

When ideal isolation point behaviour is to be achieved, although this condition is analytically sophisticated and difficult to satisfy strictly, we can derive ideal isolation point condition[29] are

$$\Delta_m = \pm J_2 \frac{\sqrt{\frac{\kappa_m}{\kappa_b}}(\gamma_d+\gamma_b-2\gamma_m\pm 2\kappa_m-\kappa_b)-\sqrt{\frac{\kappa_b}{\kappa_m}}(\gamma_d-\gamma_m\pm\kappa_m)}{(\gamma_d+\gamma_b-2\gamma_m\pm 2\kappa_m-\kappa_b)} \approx \pm 2J_2\sqrt{\frac{\kappa_m}{\kappa_b}} \quad (S13)$$

$$\delta_{\text{IIP}} = \pm J_2\sqrt{\frac{\kappa_m}{\kappa_b}} - i\kappa_m - i\gamma_m$$

The final approximation in Eq.S13 is valid for our three modes system with the (in-)equality relationship of $\gamma_b \sim \gamma_d \sim \kappa_m \ll \gamma_b \ll \kappa_b$, and thus approaches the necessary condition for IIPs and coalescence of PZR and PZT. At this point non-reciprocity is at its maximum.

In Figs. S4(a) and S4(b), two nearly-zero linewidth states (PZR points) are observed experimentally (as indicated by white arrows), $\text{Im}(\widetilde{\omega}_{ZR}) = 0$ shown in Fig. S4(h). Additionally, PZT points are observed in Figs. S4(c) and S4(d), corresponding to $\text{Im}(\widetilde{\omega}_{ZT}) = 0$ in the ZT ($S_{12} = 0$ or $S_{21} = 0$) states shown in Fig. S4h. Moreover, PZR and PZT can intersect in $\Delta_m = \pm 5 MHz$ as the Eq. S13 prophesy, corresponding to ideal isolation points in two opposite directions. Fig. S4(f) presents the real parts of the eigenvalues corresponding to each eigenstate, which agree well with the experimental observations.

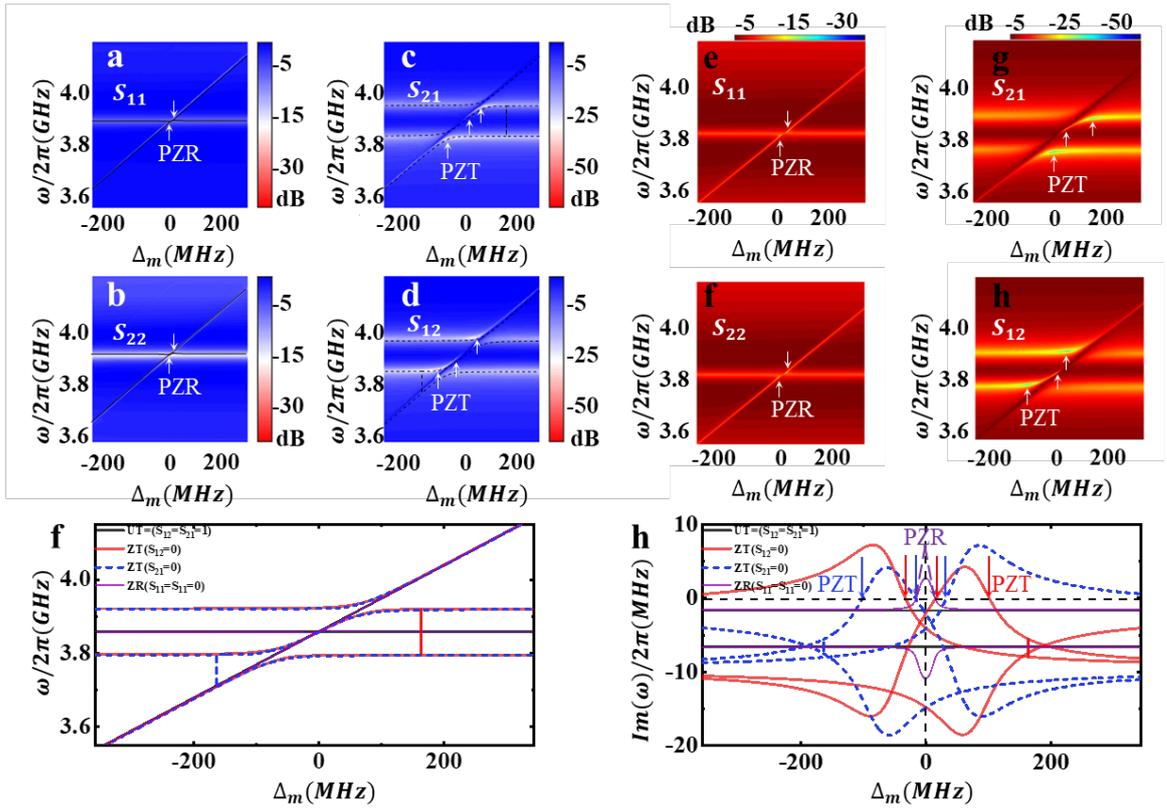

**FIG.S4 Scattering spectra and eigenvalue behaviors of the scattering Hamiltonian in Region 3.** (a)-(d) Experimental obtained spectra of four elements of the S-matrix as the function of the $\omega$ and the $\Delta_m$. (e)-(f) Calculated mapping of $|S_{11}|$ (e), $|S_{22}|$(f), $|S_{21}|$ (g), $|S_{12}|$ (h) as the function of the $\omega$ and the $\Delta_m$. The curves represent the eigenvalues of the scattering Hamiltonian. The solid (dash) black line represents ZR (ZT). ZR exhibits a behavior indicative of level attraction. Simultaneously, two PZR points appear in the reflection spectrum, and three distinct PZT points are observed in $S_{12}$ and $S_{21}$, as indicated by arrows as well as in (h). (f)-(h) The real (f) and imaginary

(h) parts of the eigenvalues of the UT(black solid line for $S_{12}= S_{21}=1$), ZT(red solid line for $S_{12}=0$, blue dash line for $S_{21}=0$),and ZR(solid purple line for $S_{11}= S_{22}=0$ ) versus $\Delta_m$ . The imaginary parts of two PZRs and PZTs coincide under the same $\Delta_m$ , resulting in an ideal isolation.

## S3 The Intersection I (zero to unit transmission (Z2U) turning point)

We aim to further investigate the impact of coherent coupling between modes on the scattering spectrum. Initially, we set an external magnetic field to adjust to $\Delta_m$. While moving the YIG crystal sphere along the Y-axis, following the centerline of the Y-waveguide, we measure its scattering parameters. The parameters at different positions are fitted using Equation S4. According to theoretical Equation S2, it is evident that $J_3$ directly influences the level repulsion of the zero-reflection state. We move the YIG sphere along the region of the dark mode distribution and observe that the spacing between the two zero-reflection points corresponds closely to the fitted $J_3$ values. As we pass through cross point one, the associated $J_3$ value suddenly drops to zero, consistent with our experimental results.

In Figure S5, experimental spectra of $S_{11}$ along paths $\ell$ (Fig. S5(a)) and paths y (Fig. S5(c)) are obtained, and the variation of the coupling parameter $J_3$ along these paths is calculated by fitting Eq. S4 (Fig. S5(b), Fig. S5(d)). Both paths show that the $J_3$ value is maximum at y=12mm. Fig. S5(d) also reflects a $J_3$ discontinuity point along the y-direction at this point, further illustrating the transition from ZT to UT due to the variation in direct coupling between the dark mode and magnon mode.

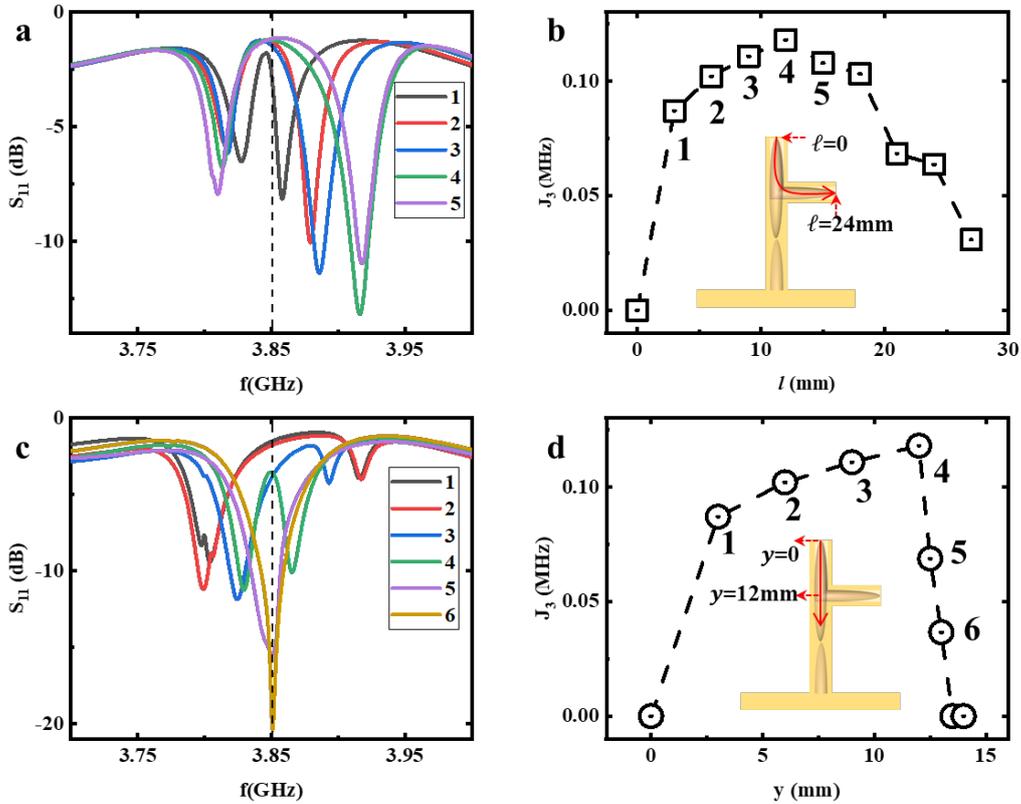

**FIG.S5 Coupling behavior in the reflection spectra at phase transition point I . (a)

Characteristics of the $S_{11}$ spectral lines at different positions when moving the YIG crystal sphere along the dark mode distribution. Panel (b) displays the corresponding fitted values of $J_3$, with different $\ell$, and the inset illustrates their coordinate representation. (c) Characteristics of the $S_{11}$ spectral lines when moving the YIG crystal sphere along the Y waveguide, passing through the phase transition point I. Panel (d) represents the corresponding fitted values of $J_3$ versus different $\ell$, and the inset illustrates their coordinate representation.

**S4 The Intersection II (non-reciprocal phase transition point)**

We further investigated the behavior when the YIG crystal sphere is at cross point **II**. Fig. S4(b) illustrates the mapping of $S_{1,2}$ as a function of $\Delta_m$ and $\omega$, while Fig. S4(d) shows the mapping of $S_{2,1}$. When the YIG sphere is placed at cross point **II**, maximum right isolation behavior occurs at blue detuning $(\Delta_m = 5MHz)$ and the maximum left isolation at red detuning $(\Delta_m = -5MHz)$. We kept $\Delta_m = 5MHz$ constant and moved the YIG sphere along the Y-waveguide. We defined the isolation degree Iso=log($|S_{21}|/|S_{12}|$) at the magnon induced resonant frequence which marked by the black raw (Fig. S4 (a)-(b)) to characterize its non-reciprocal behavior, using asymmetry Asy=log($|S_{11}|/|S_{22}|$) to describe the network's symmetry. It is noted that on the y-axis, $S_{1,1} = S_{2,2}$, indicating a symmetric network. Fig. S6a show the four Scattering spectra when the YIG is at y=36mm, representing reciprocal and symmetric network. While y=36.5mm, the Iso reach it's max (Fig. S6b). When the YIG leaves the transmission line (y=37.5mm), then the coupling effect of magnon disappears (Fig. S6c). The non-reciprocal range along the y-axis is approximately 2 mm (Fig. S6d).

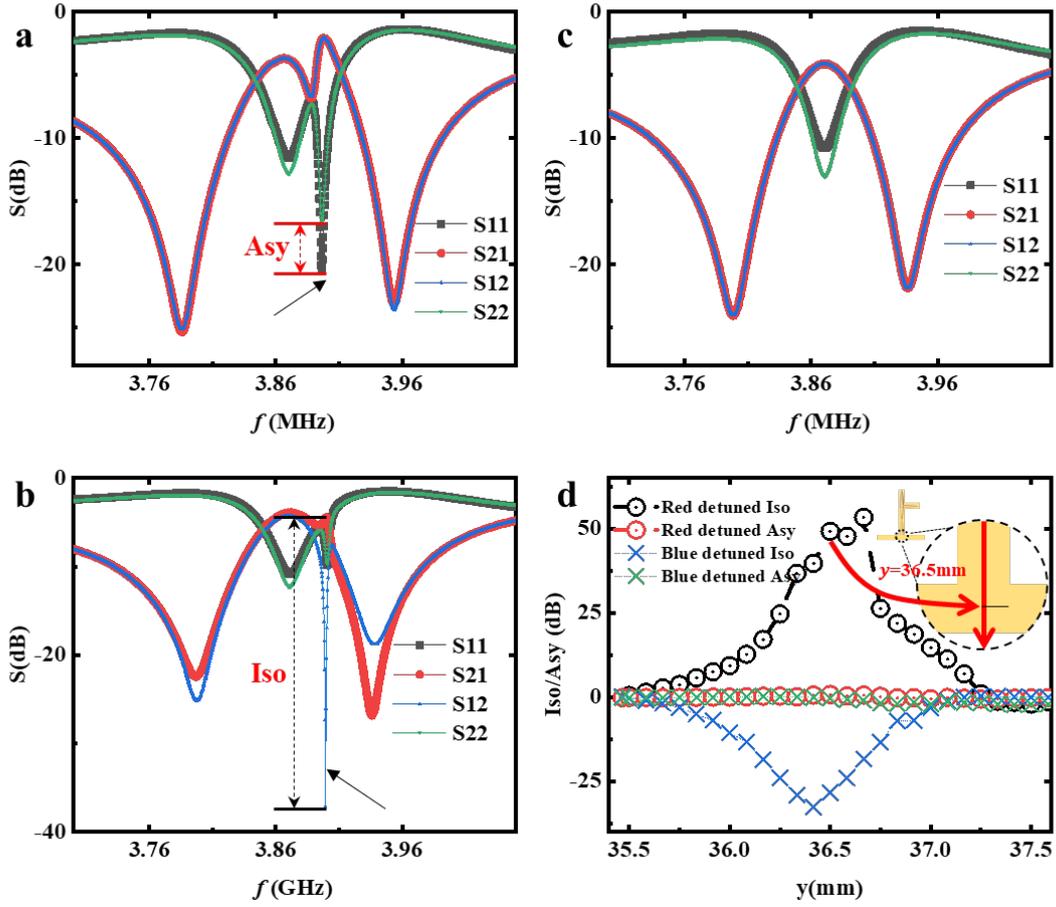

**FIG.S6 The scattering spectrum behavior at phase transition point II.** (a) A Scattering spectra behavior of sS11 (blue), s12 (red), and s21 (black) at y=36mm, representing reciprocal and symmetric network. (b) Scattering spectra behavior of sS11 (blue), s12 (red), and s21 (black) at y=36.5mm, showcasing maximum non-reciprocity. (c) Scattering spectra behavior of sS11 (blue), s12 (red), and s21 (black) at y=37.5mm, indicating the disappearance of magnon coupling features. (d) Iso and Asy variation with y position under red (blue) detuning. The width of Iso represents the width of this non-reciprocal point, which is 1mm.

For a better understanding of its non-reciprocal characteristics, moving the YIG crystal sphere along the x-axis. Figure S7 illustrates the variation of the scattering spectra as the YIG sphere moves along the X-axis. The spectra corresponding to the positions where Asy is maximum (Fig. S7(a), Fig. S7(c), Fig. S7(d), Fig. S7(f)). When the YIG crystal is positioned at x=0mm, where the isolation (Iso) is maximum, the results are presented in Fig. S7(b)($\Delta_m = -5MHz$) and Fig. S7(e)($\Delta_m = 5MHz$). Theoretical analysis indicates that Asy is maximum when propagation phase $\theta = \pm\pi$ by Eq. (S21)    and Iso is maximum when $\theta = 0$. The experimental results are in good agreement with the theoretical predictions. Figure S7(g) demonstrates that the relationship between Asy and x is centrally symmetric, whereas the relationship between Iso and x is symmetric.

The blue (red) scatterplot represents the corresponding Iso and Asy values under blue (red)

detuning $\Delta_m = 5MHz$ ($\Delta_m = -5MHz$).where the abscissa represents the position at cross point **II** on the TML.

At x=0mm, the non-reciprocity reaches its maximum at this resonant point, while the reflection remains equal. Conversely, at x=-12mm, the reflection exhibits the maximum difference, reaching the maximum asymmetric. By moving its position along the TML, we find that cross point **II** corresponds to the maximum value of Iso. It is observed that maintaining non-reciprocity spans approximately 24 mm along the x-axis.

Non-reciprocal behavior requires coherent coupling and dissipative coupling between modes. At cross point **II**, it is feasible to achieve significant values for both magnon and bright modes, thus realizing relatively ideal isolation. Regarding the emergence of reflection asymmetry, we consider x=0 as the zero-phase reference point. If the YIG crystal sphere is not on the negative half-axis (x < 0), the phase from port 1 to magnon is advanced by $\theta = kx$, while the phase from port 2 to magnon is delayed by $\theta = kx$, where $\theta = kx = 2\pi x/\lambda$, representing the electrical length. Consequently, the coupling coefficient matrix between magnon and the port should be denoted as D.

$$D = \begin{bmatrix} \sqrt{\kappa_b} & 0 & \sqrt{\kappa_m}e^{i\theta} \\ \sqrt{\kappa_b} & 0 & \sqrt{\kappa_m}e^{-i\theta} \end{bmatrix}, D^\dagger D = \begin{bmatrix} 2\kappa_b & 0 & \sqrt{\kappa_b \kappa_m}(e^{i\theta} + e^{-i\theta}) \\ 0 & 0 & 0 \\ \sqrt{\kappa_b \kappa_m}(e^{i\theta} + e^{-i\theta}) & 0 & 2\kappa_m \end{bmatrix}$$

$$H_{res} = H_0 - \frac{i}{2}D^\dagger D = \begin{bmatrix} \widetilde{\omega}_c - i\kappa_c & J_1 & J_2 - i\sqrt{\kappa_b \kappa_m}\cos\theta \\ J_1 & \widetilde{\omega}_d & 0 \\ J_2 - i\sqrt{\kappa_b \kappa_m}\cos\theta & 0 & \widetilde{\omega}_m - i\kappa_m \end{bmatrix} \quad (S14)$$

Due to the significant coupling effects caused by the propagation phase $\theta$, it is necessary to correct $H_{res}$ by Feshbach projection approach.

$$H'_{res} = H_{res} + \Delta(\omega)$$

Even if the injected microwaves are not strictly equal to the resonant frequency ($\omega$), the mode will still respond to microwaves at other frequencies. Then $\Delta(\omega)$ represents the effect of the response due to the other frequency ($\omega'$).

$$\Delta(\omega) = PV\left(\int \frac{D^\dagger(\omega')D(\omega')}{\omega - \omega'} d\omega'\right) = \begin{bmatrix} 0 & 0 & \sqrt{\kappa_b \kappa_m}\sin\theta \\ 0 & 0 & 0 \\ \sqrt{\kappa_b \kappa_m}\sin\theta & 0 & 0 \end{bmatrix} \quad (S15)$$

Then the $H'_{res} = \begin{bmatrix} \widetilde{\omega}_c - i\kappa_c & J_1 & J_2 - i\sqrt{\kappa_b \kappa_m}\cos\theta \\ J_1 & \widetilde{\omega}_d & 0 \\ J_2 - i\sqrt{\kappa_b \kappa_m}\cos\theta & 0 & \widetilde{\omega}_m - i\kappa_m \end{bmatrix} \quad (S16)$

When setting the diagonal elements of the target value of **S**-matrix $\lambda$ equal to zero, we get the effective scattering Hamiltonian for zero-reflection,

$$H_{sca}^{S_{i,i}=0} = \lim_{\epsilon \to 0} \begin{bmatrix} \widetilde{\omega}_b - i\kappa_b + \frac{i\kappa_b}{\epsilon} & J_1 & J_2 - i\sqrt{\kappa_b \kappa_m}e^{i\theta} + \frac{i\sqrt{\kappa_b \kappa_m}}{\epsilon}e^{\pm i(\theta+\pi)} \\ J_1 & \widetilde{\omega}_d & 0 \\ J_2 - i\sqrt{\kappa_b \kappa_m}e^{i\theta} + \frac{i\sqrt{\kappa_b \kappa_m}}{\epsilon}e^{\pm i\theta} & 0 & \widetilde{\omega}_m - i\kappa_m + \frac{i\kappa_m e^{\pm i(2\theta+\pi)}}{\epsilon} \end{bmatrix}$$

$$(S17)$$

zero-reflection condition is given by,

$$det\left|\widetilde{\omega}_{ZR} - H_{sca}^{S_{i,i}=0}\right| = 0. \quad (S18)$$

$$(\widetilde{\omega}_{ZR} - \widetilde{\omega}_d)[(\kappa_m - \kappa_b)\widetilde{\omega}_{ZR} + \kappa_b\widetilde{\omega}_m - \kappa_m\widetilde{\omega}_b] = \kappa_m J_1^2 \quad (S19)$$

$$(\widetilde{\omega}_d - \widetilde{\omega}_{ZR})[(\widetilde{\omega}_b - i\kappa_b - \widetilde{\omega}_{ZR})\kappa_m e^{\mp i2\theta} + (\widetilde{\omega}_m - i\kappa_\square - \widetilde{\omega}_{ZR})\kappa_b - J_2\sqrt{\kappa_b\kappa_m}(e^{\pm i\theta} + e^{\mp i\theta})]$$
$$= \kappa_m J_1^2 e^{\mp i2\theta}$$

Which is transformed into a quadratic equation of $\widetilde{\omega}_{ZR}$

$$(\kappa_m e^{\mp i2\theta} + \kappa_b)\widetilde{\omega}_{ZR}^2 - (\kappa_b\widetilde{\omega}_m + \kappa_m e^{\mp i2\theta}\widetilde{\omega}_b - (\kappa_m e^{\mp i2\theta} + \kappa_b)\widetilde{\omega}_d + J_2\sqrt{\kappa_b\kappa_m}(e^{\pm i\theta} + e^{\mp i\theta}))\widetilde{\omega}_{ZR} + \kappa_m e^{\mp i2\theta}\widetilde{\omega}_d\widetilde{\omega}_b + \kappa_b\widetilde{\omega}_d\widetilde{\omega}_m - J_1^2\kappa_m e^{\mp i2\theta} - J_2\sqrt{\kappa_b\kappa_m}(e^{\pm i\theta} + e^{\mp i\theta})\widetilde{\omega}_d = 0$$
$$(S20)$$

$$\widetilde{\omega}_{ZR\pm} = \frac{-B \pm \sqrt{B^2 - 4AC}}{2A} \quad (S21)$$

Where $A = \kappa_m e^{\mp i2\theta} + \kappa_b$, $B = -\kappa_b\widetilde{\omega}_m - \kappa_m e^{\mp i2\theta}\widetilde{\omega}_b + (\kappa_m e^{\mp i2\theta} + \kappa_b)\widetilde{\omega}_d - J_2\sqrt{\kappa_b\kappa_m}(e^{\pm i\theta} + e^{\mp i\theta})$, $C = \kappa_m e^{\mp i2\theta}\widetilde{\omega}_d\widetilde{\omega}_b + \kappa_b\widetilde{\omega}_b\widetilde{\omega}_m - J_1^2\kappa_m e^{\mp i2\theta} - J_2\sqrt{\kappa_b\kappa_m}(e^{\pm i\theta} + e^{\mp i\theta})\widetilde{\omega}_d$.

When set $\text{Im}(\widetilde{\omega}_{ZR\pm}) = 0$, we get the $\theta = \pm\pi$, which means $x = \pm\frac{\lambda\theta}{2\pi} = \pm 12mm$.

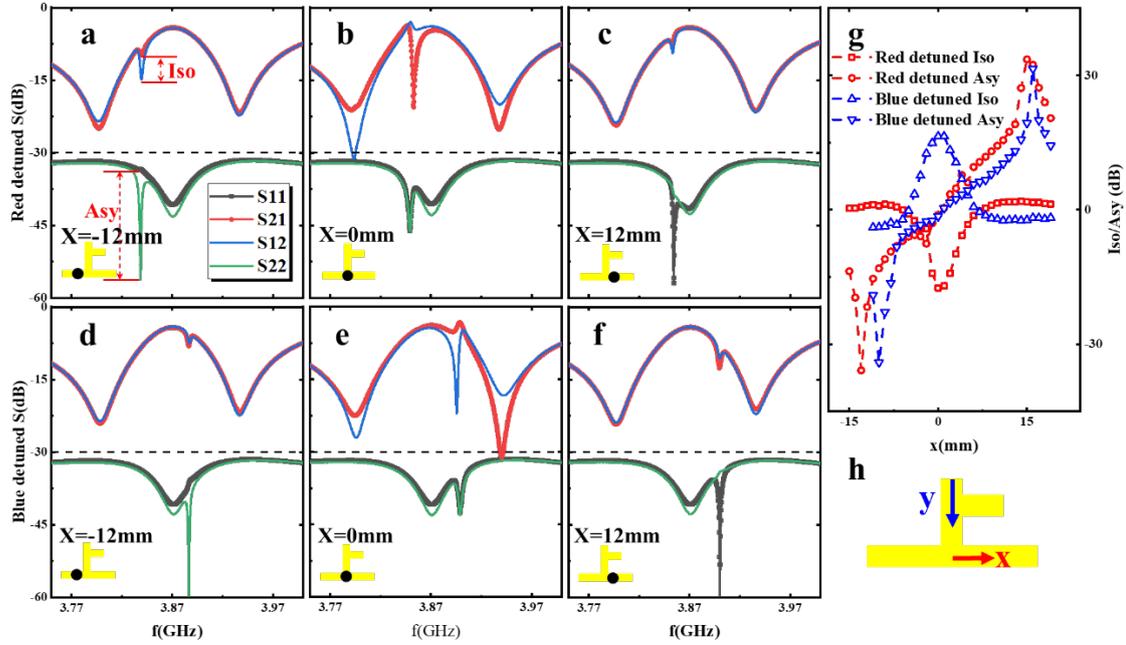

**FIG.S7 Aym and Iso results near nonreciprocal phase transition points in the system.**
(a)-(c) Scattering spectra (red (blue) curves for S12(S21), black (green) curves for S11(S22)) at the ideal isolation point under red detuning ($\Delta_m = -5MHz$), depicted through curves at different x-coordinates ((a) for x=-12mm, (b) for x=0mm, (c) for x=12mm). (d)-(f) Under blue detuning ($\Delta_m = 5MHz$), the scattering curves at different x-coordinates ((a) for x=-12mm, (b) for x=0mm, (c) for x=12mm). we define a asymmetric ratio Asy=log(S11/S22) and iso=log(S21/S12) at the magnon induced resonant frequency w=3.84 (w=3.90) for red (blue) detuning. (g) The Asy and Iso varies with the x-coordinate at the blue (red) $\Delta_m$ detuning. The Iso (Asy) is symmetrical (antisymmetrical) with respect to the X-coordinate.h The Asy and Iso varies with the x-coordinate at the blue (red) magnon detuning $\Delta_m$. The Iso (Asy) is symmetrical (antisymmetrical) with respect to the X-coordinate. (h) Diagram illustrating the coordinate axes.

| Parameters | Definition | Determination | Note | Value(MHz) |
|---|---|---|---|---|
| $\kappa_b$ | External damping rate of the bright cavity mode | Fit from the transmission spectrum when only the Y waveguide and transmission line are present | Fig. 1c | 154.7 |
| $\gamma_b$ | Intrinsic damping rate of the bright cavity mode | | | 12.65 |
| $\kappa_m$ | External damping rate of the magnon mode | Fit from the transmission spectrum when only the magnon and transmission line are present | Fig. 1e The YIG placed on the transmission line(y=36.5mm) | 1.550 |
| $\gamma_m$ | Intrinsic damping rate of the magnon mode | | | 1.714 |
| $\kappa_d$ | External damping rate of the dark cavity mode | Fit from the transmission spectrum when only Y-shaped cavity resonator | Fig. 1d | 0 |
| $\gamma_d$ | Intrinsic damping rate of the dark cavity mode | | | 6.654 |
| $J_1$ | Coherent coupling strength between the cavity bright mode and dark mode. | | | 62.57 |
| $J_2$ | Coherent coupling strength between the cavity bright mode and magnon mode. | Fit from the transmission spectrum, when placing the YIG sphere at different positions along the Y-waveguide Fig. 2c | ① y=12mm | 65.14 |
| | | | ② y=13mm | 47.98 |
| | | | ③ y=36.5mm | 22.63 |
| $J_3$ | Coherent coupling strength between the cavity dark mode and magnon mode. | | ① y=12mm | 59.00 |
| | | | ② y=13mm | 0 |
| | | | ③ y=36.5mm | 0 |
| $\Gamma$ | Dissipative coupling strength between the bright mode and the magnon mode | $\Gamma = \sqrt{\kappa_b \kappa_m}$ | y=36.5mm | 15.48 |

**Table. S1:Parameters Table**